\documentclass[conference]{IEEEtran}
\IEEEoverridecommandlockouts
\usepackage{cite}
\usepackage{amsmath,amssymb,amsfonts}
\usepackage{graphicx}
\usepackage{textcomp}
\usepackage{xcolor}
\def\BibTeX{{\rm B\kern-.05em{\sc i\kern-.025em b}\kern-.08em
    T\kern-.1667em\lower.7ex\hbox{E}\kern-.125emX}}

\usepackage{times}  %
\usepackage{helvet} %
\usepackage{courier}  %
\usepackage{graphicx} %
\usepackage{appendix}
\usepackage{makecell}
\usepackage{tikz}
\def\checkmark{\tikz\fill[scale=0.4](0,.35) -- (.25,0) -- (1,.7) -- (.25,.15) -- cycle;}

\usepackage[]{quoting}
\usepackage{adjustbox}
\usepackage{amsmath}
\usepackage{amssymb}

\usepackage{flexisym}
\usepackage{multirow}
\usepackage{comment}
\usepackage{xcolor}
\usepackage{times}
\usepackage{soul}
\usepackage{url}
\usepackage[utf8]{inputenc}
\usepackage{amsmath}
\usepackage{booktabs}
\usepackage[ruled,vlined ]{algorithm2e}
\usepackage[noend]{algpseudocode}
\usepackage{amsmath}
\usepackage{enumitem}
\usepackage{bm}
\usepackage{caption}
\usepackage{subcaption}
\usepackage{makecell}
\usepackage{xcolor}
\newcommand{\ts}{\textsuperscript}
\newcommand*{\affaddr}[1]{#1} %
\newcommand*{\affmark}[1][*]{\textsuperscript{#1}}

\newcommand{\commentout}[1]{%
}
\newcommand{\fixme}[1]{{\textcolor{red}{\textit{#1}}}}

\begin{document}
\title{Understanding Stay-at-home Attitudes \\through Framing Analysis of Tweets}

\author{Zahra Fatemi\affmark[1], Abari Bhattacharya\affmark[1], Andrew Wentzel\affmark[1], Vipul Dhariwal\affmark[1], Lauren Levine\affmark[2], \\ Andrew Rojecki\affmark[2], G. Elisabeta Marai\affmark[1], Barbara Di Eugenio\affmark[1], Elena Zheleva\affmark[1]\\
\affaddr{\affmark[1]Department of Computer Science}, \affaddr{\affmark[2]Department of Communication}\\
University of Illinois at Chicago, Chicago, IL\\
\{zfatem2, abhatt62, awentze2, vdhari3, llevin20, arojecki, gmarai, bdieugen, ezheleva\}@uic.edu
}

\maketitle

\begin{abstract}
With the onset of the COVID-19 pandemic, a number of public policy measures have been developed to curb the spread of the virus. However, little is known about the attitudes towards stay-at-home orders expressed on social media despite the fact that social media are central platforms for expressing and debating personal attitudes. 
To address this gap, we analyze the prevalence and framing of attitudes towards stay-at-home policies, as expressed on Twitter in the early months of the pandemic. We focus on three aspects of tweets: whether they contain an attitude towards stay-at-home measures, whether the attitude was for or against, and the moral justification for the attitude, if any. We collect and annotate a dataset of stay-at-home tweets and create classifiers that enable large-scale analysis of the relationship between moral frames and stay-at-home attitudes and their temporal evolution. Our findings suggest that frames of care are correlated with a supportive stance, whereas freedom and oppression signify an attitude against stay-at-home directives. There was widespread support for stay-at-home orders in the early weeks of lockdowns, followed by increased resistance toward the end of May and the beginning of June 2020. The resistance was associated with moral judgment that mapped to political divisions.
\end{abstract}
\maketitle
\section{Introduction}

The outbreak of the novel Coronavirus Disease (COVID-19) upended people’s daily lives across the globe. Emergent measures and policies, such as lockdown, mask wearing, and social distancing, were mandated by governments to prevent the spread of the virus. Citizen response to these orders varied as cooperation became entangled with political culture and partisan politics. %
Initial studies examined the topics and sentiment of attitudes towards self-quarantine mandates expressed on social media (e.g., ~\cite{Yin-adma20,li-po21}) but none have undertaken a systematic analysis of the moral justifications underlying them. %

We focus on the moral dimension of Twitter messages because attitudes buttressed by moral conviction are more predictive of behavior \cite{Bloom2013, Scott2010, Skitka2008} and are more resistant to change \cite{Aramovich2012,Hornsey2003}. Research also finds that news stories emphasizing moral judgment are more likely to be shared on social media\cite{valenzuela2017} and that people are more likely to share content that elicits moral outrage \cite{Crockett2017}. 

We focus specifically on  stay-at-home (SAH) orders, the first officially mandated measure intended to stop the spread of the virus. In some respects SAH decision-making process resembled a prisoner’s dilemma situation where cooperation, the most optimal societal strategy for reducing the spread of the virus, contended with betrayal, the most optimal outcome for individuals chafing against restrictions on their freedom. Complicating the dilemma were the political hazards to the government in power of a lockdown that would have a negative effect on the economy and thereby pose a threat to its viability. The competing incentives would play out in social media where opinion on an alarming issue would elicit moral judgment that hardened opinion that would prolong the pandemic. Based on research regarding morals and partisanship \cite{graham-jpsp09}, we expect increased support for SAH based on Care and Justice and opposition based on Loyalty, Authority, Purity, and Liberty. 

To understand users’ attitudes towards SAH, we analyze opinions expressed on Twitter through the lens of framing theory. Framing theory posits that an issue can be viewed from multiple perspectives and thereby evaluated differently depending on the view emphasized. Accordingly, the selective presentation of information can influence citizens’ judgment of issues by making some aspects more salient than others. One widely cited definition specifies that “to frame is to select some aspects of a perceived reality and make them more salient in a communicating text, in such a way as to promote a particular problem definition, causal interpretation, moral evaluation, and/or treatment recommendation for the item described”~\cite{Entman93}. As such, frames exist in two places, in the minds of citizens (frames in thought) and in media content (frames in communication)~\cite{chong2007}. The premise of framing studies is that the preponderance of media frames influence the distribution of opinion on public issues, a phenomenon widely documented on foreign and domestic issues alike 
\cite{garten-ijcai16}.

Scholars also study mediating variables that intervene between frames in communication and those in thought. Social media such as Twitter occupy an intriguing middle position in this model. This is because, although attitudes expressed on the medium represent a sample of frames in thought, Twitter may also be regarded as a medium for highlighting frames in communication. Given the profile of Twitter authors-—better educated, younger, more interested in politics \cite{wojcik-prc19}—they are also more likely to be influential within their networks of like-minded followers.

To conduct moral frame analysis and relate message frames to SAH attitudes, we collect and clean a dataset with $5.5$ million tweets posted between March and June $2020$ containing carefully selected keywords.
We annotate $7,161$ of these tweets for their relevance to SAH orders and, issue position (stance), as well as their moral framing.
Our analysis investigates the relationship between moral frames and stay-at-home attitudes, their temporal evolution, correlation, and virality. We also contrast the content characteristics of tweets originating in the United States from those abroad and examine their relationship to partisanship.
\section{Related Work}

\textbf{Attitudes towards public directives related to COVID-19.}
Social media studies have looked at citizen responses to different public directives related to COVID-19.
Li et al. \cite{li-po21} explored public supporting and opposing perceptions towards reopening policy from
 both temporal and spatial perspectives and found that online perceptions could vary in the appearance of influential news and could be associated with demographic and socio-economic characteristics. 
Other studies focused on the association between political preferences and following governors’ recommendations for individuals to stay at home and found that there is a significant gap in the people's beliefs about social distancing and COVID risks.
~\cite{grossman-nac20,allcott-jpe20}.
Others studied whether the political affiliation affects compliance with social distancing orders, observing that Republicans were less likely to follow social distancing orders than Democrats~\cite{painter-jebo21}. Studies also explored beliefs influenced misinformation about COVID-19 and the role of such misinformation on influencing compliance~\cite{roozenbeek-jrc20}, including those related to vaccination %
~\cite{johnson-nature20,wen-ying-hc20}.
Sentiment analysis of COVID-19 Twitter posts found that most of the "stay safe home" tweets had positive sentiments~\cite{Yin-adma20}. Topic analysis of anti‐quarantine comments revealed that they were related to mask wearing and political issues such as the impact on elections~\cite{karami-aist20}. In contrast to previous work, we focus on the framing aspects of stay-at-home attitudes.

\textbf{Measuring attitudes in social media.} 
Over the last decade, there has been active research in stance detection in tweets for understanding attitudes, covered by a recent survey~\cite{aLdayel-ipm21}. Sobhani et al. proposed a stance detection system and showed that even though sentiment features are useful for stance detection, they alone are not sufficient \cite{sobhani-lcs16}.
Darwish et al. developed an
unsupervised framework for detecting the stance of Twitter users with respect to controversial topics \cite{darwish-icwsm20}.
A SemEval stance dataset has been created with tweets labeled with sentiment and stance as ground truth to help evaluate the new stance detection algorithms~\cite{mohammad-semeval16}. 
Using the Linguistic Inquiry and Word Count (LIWC) tool and Meaning Extraction Method (MEM), Mitra et al. %
characterize the long-term advocates of pro- and anti-vaccination attitudes \cite{mitra-icwsm17}.
Hassan et al. developed a random walk method to identify the attitude of participants in an online social network discussion toward one another ~\cite{hassan-emnlp10}.

\textbf{Moral frame analysis.}
Moral frames correlate with the political affiliation of people \cite{graham-jpsp09} and news stories containing a moral frame are more likely to be shared on social media \cite{valenzuela-joc17}, especially so on emotionally charged issues such as health. 
Prior studies identified moral frames in news articles \cite{card-emnlp16}. Shahid et al. have found that
moral frames have a skewed distribution with Cheating and Harm being the dominant frames \cite{shahid-acl20}. Scholars have also conducted
Frame prediction of Twitter data using collective classification \cite{johnson-acl17} and distributed representations \cite{garten-ijcai16}.
Identifying moral frames in tweets requires considerable domain expertise and training.
The  Moral Foundations Twitter Corpus was developed to facilitate this process~\cite{hoover-spps20}.
Kaur et al. ~\cite{kaur-bigdata16}
propose an approach to quantify the relationship between moral frames and conversations around abortion, homosexuality, immigration, religion, and immorality topics on Twitter and observed that 
 Care is most dominant foundation and Purity is the most distinctive foundation in conversations on immorality. %

\section{Data harvesting and annotation}

\label{sec:data}
Collecting a representative sample of tweets is fraught with difficulties, because of the sheer number of tweets, the non-standard  language, and the amount of noise  \cite{eisenstein-naacl2013,sarker-norm2017}. As commonly done \cite{skaik-acm17}, we  start with keywords, but take several steps to ensure the quality of our data.

First, we selected the 87M  tweet-IDs between March $1$ and June $30$ $2020$ from ~\cite{chen-jmir20}, %
one of the largest publicly available tweet datasets with COVID-19-related keywords.
We used Twarc~\cite{morstatter-aaai13} to collect the corresponding content (text, hashtags, etc) and authors’ metadata.

Among those 87M tweets,  we selected 42.3M  English tweets  (tweet attribute $lang$). Then, we used topic analysis to find selective keywords.
\commentout{In this section, we describe how we select and annotate our final dataset for in-depth analysis.}

\noindent
\textbf{Finding underlying topics.} We remove URLs, %
make all words lowercase, remove punctuation and stopwords, and lemmatize and stem the words. \commentout{We use topic analysis to understand the different topics of COVID-19 tweets and identify the main keywords for the stay-at-home topic.} 
We apply
Latent Dirichlet Allocation (LDA) \cite{blei-jmlr} to a $5\%$ random sample of the COVID-19 English tweets. %
We
extract $20$ topics (chosen based on coherence), with the corresponding words and their probabilities.
\commentout{words associated with the topics, and probability of each word being represented in a topic. } \commentout{Since this step was done early in the analysis, it included a $5\%$ random sample of the COVID-19 English tweets from March to May only.} By manual inspection and consensus between two of the authors, we agreed that four of the discovered topics appear to  relate to stay-at-home. Table~\ref{table:lds_keywords} shows the top $10$ words for each of these four topics.
\begin{table}[h]
\vspace{-3pt}
\centering
\small
\setlength{\tabcolsep}{5pt}
\begin{tabular}{|c|c|c|}
 \hline
Topic&  Word   \\ \hline
1    & \makecell{ home,  stay, corona, virus, people, work, love, \\stop, want, stick}    \\ \hline
2    & \makecell{quarantin, covid, coronavirus, open, stayhom, \\like, thank,  pandem, look, quarantinelif}    \\ \hline
3    & \makecell{ home, time, stay, distanc, social,\\ covid,  life, miss, stayhom, f**k}    \\ \hline
4    & \makecell{home, stay, order, peopl, work,  need,\\close, open, essenti, coronavirus}    \\ \hline
\end{tabular}
\caption{Ten top words in selected LDA topics%
}
\label{table:lds_keywords}
\vspace{-10pt}
\end{table}

\noindent
\textbf{Finding stay-at-home candidate keywords}: %
The  top $10$ words in the four LDA topics and  their synonyms from WordNet~\cite{miller-princeton} result in 
$60$ candidate  keywords. 
\commentout{as candidate keywords to help us identify tweets concerning stay-at-home. %
Using WordNet~\cite{miller-princeton}, we identify synonyms of the candidate keywords, for a total of $60$ candidate stay-at-home keywords.}
For each keyword, we sampled $100$ tweets containing the keyword and manually checked if at least $80$ of the tweets concern  stay-at-home. If so,  the keyword is included in the final five keyword set:  \emph{home, open, quarantine, lockdown, inside}. This was motivated by seeking a low false-positive rate for keyword-based filtering. \commentout{The final set of stay-at-home keywords is \emph{home, open, quarantine, lockdown, inside}. } 
Finally, we filter the \commentout{English} dataset for tweets containing at least one of the five stay-at-home keywords. %
The resulting dataset, which we name  $SAH$ corpus, %
contains $5,538,993$ tweets, excluding retweets.
\commentout{, temporally distributed as follows: %
$487,471$  from March, $850,128$  from April, $1,165,906$ from May, and $3,035,488$  from June. Additionally,}  Given our interest in both global and local (i.e., in the USA) attitudes, we %
also identify the tweets in $SAH$ that are geotagged in the U.S. based on $country\_code=US$, resulting in  $36,744$ tweets. %
\commentout{\fixme{BDE: is the next an important comment?}
Note that we do not filter out for bots, because bot tweets contribute to the ecosystem of expressed attitudes that can influence opinions online.}

\subsection{Identifying potentially relevant tweets}

\label{sec:relevance1}

The 5.5M  SAH corpus %
is  relevant to stay-at-home by construction, however we are interested in a more specific notion of relevance, i.e., attitudes  for or against stay-at-home directives.
\commentout{We distinguish between tweets from the general stay-at-home topic, defined as opinions and experiences of a private person relating to staying at home by themselves or others during the COVID-19 pandemic, vs. tweets that are relevant for our research, i.e. tweets %
whose content expresses support for or against public stay-at-home directives.
Since we are interested in tweets that are relevant specifically to attitudes towards stay-at-home directives, we further filter the dataset using a binary classifier.} 

\commentout{\fixme{BDE: was this initial relevance annotation also done phrase by phrase?}

\fixme{ZF->BDE: The file that has been shared with me contains full text of tweets. It seems that the full text was considered for the annotation.}}
Framing experts on our team annotated a random sample of $5,013$ tweets from the %
SAH corpus ($0.1\%$ of each of March, April, May) and identified $856$ tweets as relevant (based on the full text of the tweet). %
We then    trained  an SVM  relevance  classifier to aid in corpus development (input: tweet text converted to   $word2vec$ vectors~\cite{mikolov-acm}; 5 fold cross-validation; optimized hyperparameters  \{kernel = 'rbf', C = 10, $\gamma$ = 0.1\}).
\commentout{To do so, we converted the %
tweet text to vectors using the $word2vec$~\cite{mikolov-acm} model, pre-trained on Google news,  and average the vectors into a single vector for each tweet. We use 5-fold cross validation to train a Support-Vector Machine (SVM) model.
The optimized hyperparameters for the model are \{kernel = 'rbf', C = 10, gamma = 0.1\}. The model results in $86\%$ accuracy and $69\%$ precision.} This classifier ($86\%$ accuracy and $69\%$ precision)  predicts $7,161$  potentially relevant tweets from the U.S. geotagged data. We name this set of $7,161$ tweets the \emph{US-SAH} corpus.

\subsection{The US-SAH corpus:  Relevance, moral frame and stance annotation}

\commentout{We initially focused on the \emph{U.S. dataset} and later expanded our analysis to include all tweets with stay-at-home attitudes.} 
The framing experts on our team manually annotated the %
\emph{US-SAH} tweets
for (actual) relevance, moral frames and stance towards SAH. All annotations were conducted at the phrase level, since   tweets may consist of multiple phrases, each of which  may express  one moral frame.  Phrases are  grammatical units (from  bare nouns to  complex sentences) or hashtags which contain at least one  moral frame  keyword as defined in the Moral  Foundation dictionary~\cite{graham-jpsp09}, augmented as discussed below.  
Our coding protocol (with categories, values, definitions, and examples) is summarized in  Table~\ref{table:protocol}. %

$2,400$  phrases were annotated as relevant, and $27$ as somewhat relevant (Cohen's $\kappa=0.744$, computed on 7\% of the data). This resulted in $1,779$ relevant tweets (they contain at least one relevant phrase); $26$ somewhat relevant tweets (they contain only somewhat relevant phrases); and $5,356$ irrelevant tweets (otherwise). 
\commentout{In turn,  a tweet is considered {\it relevant} if it contains at least one relevant phrase;
{\it somewhat relevant} if it only contains {\it somewhat relevant} phrases; {\it irrelevant} otherwise. }
\commentout{If the tweet contains phrases that are tangentially related, wherein the attitude expression is inferred or not clearly stated, it is  considered somewhat relevant. If the tweet does not contain any phrases directly or tangentially related to stay-at-home, it is  considered irrelevant.}

Only the $1,779$ relevant tweets were annotated for moral frame and stance (at the phrase level). We name this annotated set \emph{US-SAH-MF}. Stance has  three possible values, {\it Pro/Against/Undecided}; intercoder agreement was $\kappa=0.804$ (on the same 7\% of data). 
\commentout{We used the moral foundation dictionary created by Graham et al. \cite{graham-jpsp09}, and used among others by \cite{hoover-spps20}. } For moral frames, we started with the dictionary from \cite{graham-jpsp09}, which contains words and word stems associated with ten moral frames. We added {\it Freedom} and {\it Oppression}, based on Liberty, the sixth foundation of MFT suggested by \cite{haidt2012}. For these two frames,  we 
identified 22 keywords or stems such as {\it  autonomy,  democracy,   flexib*, liberat*, open,  self-determ*}.

A random sample of $15\%$ of the tweets were doubly annotated for moral frames with high  intercoder agreement, $\alpha =0.804$ \cite{krippen80}.

\commentout{
\begin{table}
\begin{tabular}{|l|l|r|} \hline
Phrase & Moral Frame & Stance \\ \hline
{\it \small Do your part.} & Loyalty & +\\
{\it \small Stay home = stay healthy.} & Care & + \\ \hline
\end{tabular}
\caption{Annotation for the tweet {\it Do your part. Stay home = stay healthy. Its really that simple!} } %
\label{table:ex-tweet}
\end{table}
}

\begin{table*}
\centering
\caption{The protocol deployed by domain experts to annotate the datasets, together with example tweets. }
\label{table:protocol}
\small\addtolength{\tabcolsep}{-2pt}
\begin{tabular}{|p{1.4cm}|p{1.6cm}|p{5.4cm}|p{8cm}|}
\hline
 Category& Value & Definition &Example tweet\\
 \hline
    \multirow{3}{*}{Relevance}&Relevant&Tweets clearly expressing an attitude related to Stay at Home directives.&Same thing you do when you are sick stay home if you dont feel good stay home plain and simple!! \\ 
    \cline{2-4}
    &\thead{Somewhat \\relevant}&
    Tweets tangentially related to Stay at Home and/or the attitude expression can be inferred but is not clearly stated.
    &"So if you protest police, you won't catch Coronavirus, but if you protest stay-at-home orders, you will."\\
    \cline{2-4}
    &Irrelevant&
    Tweets not related to Stay at Home&Mean girls playing at the drive in. Should I go or stay home\\
    \hline
    
    \multirow{3}{*}{Stance}&Pro&Relevant tweets with attitudes in support of Stay at Home orders.&This is what irritates me. STAY HOME! Its so unfair to others. \\ 
    \cline{2-4}
    &Against&Relevant tweets in opposition of Stay at Home or in support of reopening.
    &Were doing exactly what the people who created coronavirus wants us to do. Go in the house pass it to your families and slowly die in the comfort of your own home. \#CommonSense\\
    \cline{2-4}
    &Undecided&Relevant tweets with an attitude that is vague or contains a conflicting message.&Obama wants you home\\
    \hline
    
    \multirow{12}{*}{\thead{Moral \\Foundation \\Frame}}&Care& Moral reasoning based on the need to help or protect oneself or others.& Thats why its important to stay the fuck home, whomever is able to do so. We MUST slow down this spread to reduce suffering and save lives. \\
    \cline{2-4}
    &Harm& Moral reasoning based on the fear of damage or destruction to oneself or others.
    & \#boston \#cambridge I implore you to consider the people of your cities over the economy. Many more of us will have to go back to work upon new reopening phases, and its clearly not going to be safe, no matter the costly precautions.
    \\
    \cline{2-4}
    &Loyalty& Moral reasoning based on the needs of the collective or group allegiance.&
     I stay home because it is the right thing to do for my community \#whyistayhome \#StopCOVID
    \\
    \cline{2-4}
    &Betrayal&Moral reasoning based on the judgement of unfaithfulness or acting against the needs of the collective.&People who stay open when we should ALL be closed are the reason this pandemic continues to spread.
    \\
    \cline{2-4}
    &Authority& Moral reasoning based on respect for authority figures or rules.&Correct! Italy didn't quarantine people fast enough. China is authoritarian.  When you say go home and stay there, people do what they are told. We are on Italy's path.
    \\
    \cline{2-4}
    &Subversion& Moral reasoning based on negative judgment of the rebellion against authority figures or rules..&Stay home. Don't listen to POTUS. It's real folks.
    \\
    \cline{2-4}
    &Purity& Moral reasoning based on piety or the fulfillment of religious obligations.&bro im sure god will forgive u if u stay home from church for a few weeks to protect yourself and others from a virus
    \\
    \cline{2-4}
    &Degradation& Moral reasoning based on negative judgment of  depravity or failure to fulfill
religious obligations.&stay home u damn heathens
    \\
    \cline{2-4}
    &Fairness& Moral reasoning based on the need for justice or equality.&  These celebrities act like many parts of the country arent under stay-at-home orders.  If the rest of us regular citizens have to be obedient to the law and stay at home, celebrities should be no different.
    \\
    \cline{2-4}
    &Injustice& Moral reasoning based on the fear of prejudice, inequality, or wrongdoing.&This is what irritates me. STAY HOME! Its so unfair to others.
    \\
    \cline{2-4}
    &\thead{Freedom}& Moral reasoning based on the need for freedom or constitutional rights.
 &There's not a chance in hell that the government is going to tell me to stay in my house if i want to go out. We still live in a free country.( for now anyway) \#ThinTheHeard \#SaferAtHome.
    \\
    \cline{2-4}
    &Oppression& Moral reasoning based on the fear of tyranny, subjugation, or loss of constitutional rights.& You are a Governor, not the Monarch of Minnesota! Lets put this lockdown to a Democratic vote. Our country is a Republic, not a Monarchy
    \\
\hline
\end{tabular}
\end{table*}

Table~\ref{table:us-sah-mf-stance} shows the distribution of moral frames and stance values across the $2,427$ phrases, across the~$1,805$ relevant \& somewhat relevant tweets. 
\begin{table}
\centering
\setlength{\tabcolsep}{5pt}
\small
\begin{tabular}{|l|r|r|r|r|}
\hline
Moral & \multicolumn{3}{c|}{Stance} & Total  \\ \cline{2-4}
Frame & For & Against & Undecided & \\\hline
Care & 1049 & 16 & 10 & 1075 \\
Harm & 420 & 93 & 10 &  523 \\
Fairness & 17 & 1 & 1 & 19 \\
Injustice & 16 & 11 & 0 & 27 \\
Loyalty & 232 & 9 & 3 & 244 \\
Betrayal & 49 & 5 & 1 & 55 \\
Authority &  174 & 5 & 1 & 180\\
Subversion & 25 & 28 & 11 & 64\\
Purity & 40 & 3 & 0 & 43 \\
Degradation & 6 & 1 & 0 & 7 \\
Freedom & 26 & 94 & 6 & 126 \\
Oppression & 4 &  60 & 0 & 64 \\
\hline
Total & 2058 & 326 & 43 & 2427\\
\hline
\end{tabular}
\caption{Moral frames by stance (phrase annotation)}
\label{table:us-sah-mf-stance}
\end{table}
Taking tweets as units of analysis, $1,348$ are annotated with one moral frame, $345$ with two, $75$ with three, $28$ with $4$, $5$ with $5$, and $3$ with $6$. However, the multiple moral frames that a tweet is annotated with are often not  distinct; only $81$ are annotated with moral frames which are opposite, such as {\it Care/Harm}. 
\commentout{of the $345$ annotated with two moral frames, $68$ are annotated with the same moral frame;  of the 111 annotated with at least 3 moral frames,  only 15 remain annotated with 3 distinct moral frames, 9 with four, and one with 6. } 

Even if stance was  annotated at the phrase level,  all stances within a single tweet are the same.
$1,484$ tweets ($83\%$) are in favor of SAH;  $265$  ($15\%$) are against; and only $30$ ($2\%$) are undecided. 

Additionally, we use \textit{Valence Aware Dictionary for Sentiment Reasoning (VADER)}~\cite{hutto-aaai14} to find the sentiment of tweets. VADER is a lexicon and rule-based sentiment analysis tool for social media that measures negative, positive, neutral and compound (aggregated score) sentiment scores and requires no training data. %
We set the compound cutoff score to be $0.25$ so that tweets with scores higher than $0.25$ are considered as positive, tweets with scores lower than $-0.25$ are considered as negative, and the rest are neutral. %

\section{The SAH Corpus: Automatic Labeling} 

\commentout{A slightly reduced  \emph{US-SAH} corpus with the goal of automatically annotating and analyzing the much larger \emph{SAH corpus}.} 
To label the much larger {\it SAH} corpus, we build classifiers on the \emph{US-SAH} corpus for relevance classification and on the \emph{US-SAH-MF} corpus for stance and moral frame classification. We utilize 5-fold train/test splits, and assess the performance of each classifier after hyperparameter tuning. The best classifier %
is then retrained on the whole annotated %
corpus and applied to the {\it SAH} corpus, as described in Section \ref{global}. The machine learning models we experimented with are a mix of  traditional algorithms such as SVM and Random Forest, and contemporary ones such as BiLSTM. Our choices were dictated by the small amount of annotated data,  \commentout{by the need of maintaining some explainability,} and by the need of quickly annotating the large $SAH$ corpus. Additionally, for moral frame classification,  Snorkel \cite{snorkel} combines  user defined heuristic functions, and trained classifiers.

\subsection{Relevance and stance classifiers}

\commentout{We remove  the 26 {\tt somewhat relevant}  tweets from the \emph{U.S. dataset} %
and train a new SVM relevance classifier on the remaining $7,109$ tweets. }
The  first classifier for relevance we  described in Section~\ref{sec:data}, was based on a more holistic notion of relevance, not  on phrases. Hence, we train several classifiers (SVM, Random Forest, and LSTM) on the {\it US-SAH} corpus, with  $word2vec$ embeddings as tweet features. SVM results in the best performance  ($78.7\%$ accuracy and $63\%$ precision, optimal parameters \{kernel='rbf',C=1 , $\gamma=1$\}.  \commentout{Similar to the initial relevance classifier described in Section~\ref{sec-data}, we use the $word2vec$ embeddings as tweet features.  The optimal specification for the SVM model  is \{kernel='rbf',C=1 ,gamma=1\}. The model results in $78.7\%$ accuracy and $63\%$ precision. }

For stance, we use the $1,779$ tweets in \emph{US-SAH-MF} directly, and not the phrases, because all phrases within any given tweet have the same stance. %
Oversampling is done on each training dataset of the train/test splits using the ADASYN \cite{adasyn} Python library.  To represent words in the vocabulary,  100d pre-trained GloVe  word embeddings  trained on Twitter data \cite{pennington-etal-2014-glove}  are used. Analogously, we trained SVM, Random Forest and BiLSTM 
with an Embedding layer, a Bidirectional LSTM layer and a 3 unit dense layer (please see the appendix for the  optimal hyperparameters for each of these classifiers).
\commentout{
The best hyperparameters for the model have 128 hidden units and a dropout of 0.2 in the Bidirectional LSTM layer, softmax activation, categorical cross-entropy loss function and Adam optimizer. The model is trained for 50 epochs with early stopping. }
Table \ref{table:stance_classf} shows results on stance, in terms of weighted F-scores; the BiLSTM classifier outperforms the other models.
\begin{table}
\centering
\small
\begin{tabular}{|c|c|}\hline
Models &  \makecell{Weighted F-score} \\\hline
Random (baseline) &  0.43 \\
 SVM & 0.75 \\
 Random Forest  & 0.75 \\
 BiLSTM  & \textbf{0.78} \\
\hline
\end{tabular}
\caption{Stance Classification}
\label{table:stance_classf}
\vspace{-5pt}
\end{table}

\subsection{Moral frame classifier}
\label{moral_frame_c}

\commentout{
trained classifiers along the same lines we have discussed so far, but then  we combine the best with  Snorkel's generative model takes the output of labeling functions and either assigns a probabilistic label or abstains from labeling a data point.} 

Since the dataset is small ($2,400$ phrases)  and the class distribution is unbalanced (see Table~\ref{table:us-sah-mf-stance}), %
we undersample or oversample  certain frames in each training set of the five train/test splits, in order to have $500$ instances of each frame. Specifically, \textit{Care} is undersampled;  %
we add tweets from the MFTC dataset \cite{hoover-spps20}  for  the other  moral frames.\footnote{Since MFTC is annotated for moral frames at the tweet level, we add the whole tweet, as a single phrase.}
\commentout{\textit{Harm, Fairness, Injustice, Loyalty, Betrayal, Purity, Degradation, Authority and Subversion}.} 
Since \textit{Freedom} and {\it Oppression}  do not appear in MFTC, we oversampled them from our data with  ADASYN. %

We train the same three models: SVM, Random Forest, and a BiLSTM with an Embedding layer, a Bidirectional LSTM layer  and a 12 unit dense layer (please see the appendix for optimal hyperparameter values). The same 100d pre-trained Glove embeddings are used (other features, such as POS and sentiment  by VADER~\cite{hutto-aaai14}, were experimented with, but with worse results). %
\commentout{with Python libraries (scikit-learn for the first two, Keras for the third):   SVM  %
(with optimal hyperparameters \{C = 10, gamma = 0.01, kernel = rbf\}, Random Forest %
(with optimal hyperparameters  \{n\_estimators = 100, criterion = entropy\} and a \textit{BiLSTM}.}

Table~\ref{table:moralframe_classf} presents results for three baselines.
A random classifier  randomly assigns frames to the phrases. A keyword classifier assigns the frame associated by the MF dictionary with the keyword  contained  in the phrase; %
if  more than one applies,  the one matching ground truth (if any) is considered a true positive. 
The last baseline is a Random Forest classifier trained on the MFTC. %

\begin{table}[h]
\centering
\small
\setlength{\tabcolsep}{5pt}
\begin{tabular}{|c|c|}\hline
Models & \makecell{Weighted  F1} \\\hline
 Random  & 0.09 \\
 Keywords  & 0.34 \\
 Random Forest  (MFTC) & 0.28\\
 SVM  & 0.39 \\
 Random Forest  & 0.40 \\
 BiLSTM & \textbf{0.64}\\ \hline
\end{tabular}
\caption{Moral Frame Classification}
\label{table:moralframe_classf}
\vspace{-12pt}
\end{table}
The results obtained by the BiLSTM  on the \emph{US-SAH-MF} corpus (F1=0.64)  (last line in Table~\ref{table:moralframe_classf}) are  usable.  While they are lower than the reported best performance \commentout{(via LSTM)} across the MFTC corpus, F1=0.8 \cite{hoover-spps20}, performance on some subcorpora in MFTC is lower  (Baltimore, F1=0.69; Davidson, a very low F1=0.14). Indeed,  our Random Forest baseline trained on MFTC and applied to our data performs very poorly, with  F1=0.28, suggesting that our \emph{US-SAH-MF} corpus is substantially different from MFTC.

\subsection{Labeling the SAH corpus} 

\label{global}

For each of the relevance and stance models, we take the best classifier with the optimal parameters, %
retrain the models on the complete \emph{US-SAH} corpus for relevance and \emph{US-SAH-MF} for stance and moral frames, and then use them to label the unlabelled portion of the $SAH$ corpus (the \emph{US-SAH-MF} $1,779$ relevant tweets are included in the following counts, but the gold standard labels are retained). For relevance, the final SVM classifier labeled $206,333$ out of $5,538,993$ tweets as relevant. We apply the  \textit{BiLSTM} classifier for stance to these $206,333$ relevant tweets to which we refer as \emph{SAH-REL}. %
For moral frames,  we  explore   Snorkel \cite{snorkel}, which can  combine human expertise  with the patterns uncovered by machine learning.
Snorkel  is  a weakly supervised approach which can integrate  into a generative model,  user defined labeling functions (e.g.  heuristics), and models trained on the  data.
For our task, we define $13$ labeling functions. The first $12$  assign  the moral frame associated with the keywords from the MF dictionary~\cite{Graham2013}.
The 13th labeling function is the   best  classifier (BiLSTM) from the cross-validation experiments, retrained 
 with the optimal hyperparameters from cross-validation on the entire dataset of $30,000$ phrases from the under-/over-sampled training folds. 
For each phrase, Snorkel produces a label or abstains from labeling, in which case, we apply the same BiLSTM model just described in a pipeline fashion. 
Snorkel labels $65\%$ of the $206,333$ relevant tweets with one of the $12$ moral frames and abstains from %
the remaining $35\%$, which are labelled by 
the \textit{BiLSTM} classifier.
To check whether Snorkel adds value to the pipeline, a framing expert on our team manually annotated a small random sample of the \emph{SAH-REL} dataset. Snorkel with BiLSTM outperformed BiLSTM (weighted-F1 of 0.52 vs. 0.42). %

The final distribution of  moral frames and stances is shown in Table~\ref{table:mf-stance}. %
The same general trends appear in Tables~\ref{table:us-sah-mf-stance} and~\ref{table:mf-stance}, but somewhat mitigated: e.g., $83\%$ of tweets in \emph{US-SAH-MF} are for SAH, but in the \emph{SAH-REL} corpus, only $69\%$ are. Among moral frames, although {\it Care} is still disproportionately represented, it decreases from $44\%$ of the data in \emph{US-SAH-MF}, to $40\%$ in \emph{SAH-REL}. 

Note that we do not filter out for bots, because bot tweets contribute to the ecosystem of expressed attitudes that can influence opinions online. A check for duplicate tweets revealed that $99.2\%$ of the \emph{SAH-REL} tweets are unique, indicating a low level of tweet replication by bots.

\commentout{
\{\textit{Care}: $82,530$; \textit{Harm}: $24,611$; \textit{Freedom}: $22,681$; \textit{Oppression}: $3,842$; \textit{Fairness}: $1,653$; \textit{Injustice}: $1,684$; \textit{Loyalty}: $36,449$; \textit{Betrayal}: $2,746$; \textit{Purity}: $5,128$; \textit{Degradation}: $6,780$; \textit{Authority}: $16,908$; \textit{Subversion}: $1321$ \}.}

\section{Characterizing stay-at-home attitudes on Twitter}

We begin our analysis with the association of moral frames with SAH attitudes in our relevant \emph{SAH-REL} dataset (N=206,333). As we hypothesized, a moral foundation correlated with liberal ideology--Care--has the highest proportion of tweets in support of SAH ($82.7\%$), but we also find that Subversion supports the second highest fraction ($72.1\%$). Although the latter foundation is associated with conservative political views, a closer look at the substance of a greater fraction of the tweets shows that posters assert the authority of local officials as well as medical experts as they criticize those who violate SAH orders. Also in line with our first hypothesis, moral  foundations correlated with conservative political preferences--Freedom and Oppression-had the lowest proportions of supporting SAH tweets, with $60.9\%$ and $60.8\%$, respectively. We also find that Injustice offers the least support for SAH ($58.2\%$). The comparatively smaller base of support for SAH directives referencing Injustice is based largely on messages that complain about those who violate SAH orders.

\begin{table}
\centering
\caption{\emph{SAH-REL} corpus: moral frames by stance (with $\chi^2$ residuals; boldface vs. underline: association with positive vs negative stance).}
\label{table:mf-stance}
\setlength{\tabcolsep}{2pt}
\begin{tabular}{|l|r|r|r|r|}
\hline
Moral Frame & \multicolumn{3}{c|}{Stance} & Total  \\ \cline{2-4}
& Positive & Negative & Undecided & \\\hline
{\bf Care} & 64906 (75.89) & 13554 (-77.69) &4070 (-5.14) & 82530 \\
\underline{Harm} &15089 (-28.56) & 8120 (28.48)  &1402 (3.42) & 24611 \\
Loyalty & 24501 (-9.01) &10269 (12.60) &1679 (-5.99) & 36449 \\
Betrayal & 1860(-1.67) & 816 (5.02) & 70 (-6.37) & 2746 \\
Authority &11329 (-6.45) &4651 (6.06) & 928 (1.51) & 16908\\
Subversion & 896 (-1.08) & 345 (0.46) & 80 (1.33) & 1321\\
\underline{Purity} & 3044 (-15.45) & 1876 (18.32)  & 208 (-3.85) & 5128 \\
Degradation & 407 (-12.96) & 2187 (12.85) & 386 (1.70)  & 6780 \\
Fairness & 972 (-9.19) & 596(9.82) & 85(-0.18) & 1653 \\
Injustice &947 (-11.57) & 681 (14.05) & 56 (-3.54) & 1684 \\
\underline{Freedom} & 12778 (-44.47) &8197 (38.72) &1706 (16.34)& 22681 \\
\underline{Oppression} & 2251 (-14.38)& 1448 (17.40) & 143 (-4.26) & 3842 \\
\hline
Total & 142780  & 52740 & 10813 & 206333\\
\hline
\end{tabular}
\vspace{-10pt}
\end{table}

\begin{figure*}
\centering
\includegraphics[width=\textwidth]{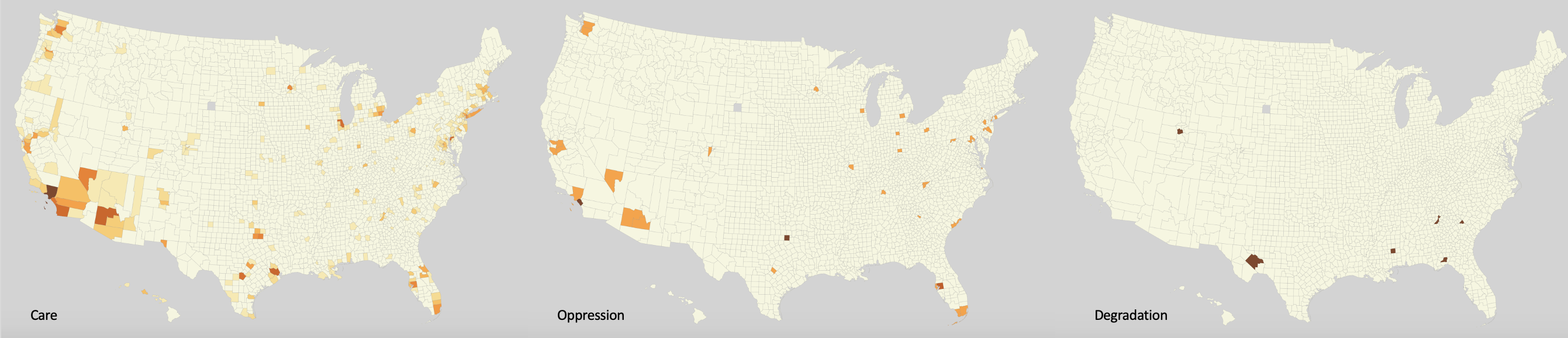}
        \caption{Geospatial distribution of tweets reflecting respectively the Care, Oppression, and Degradation moral frames. Darker shades indicate higher tweet counts for that frame~\cite{trelles2019visual}. Both Oppression and Degradation are highly localized geographically. %
        }
        \label{fig:geothreeframes}
        \vspace{-10pt}
\end{figure*}
\begin{figure*}
    \centering
        \includegraphics[width=\textwidth, height=210pt]{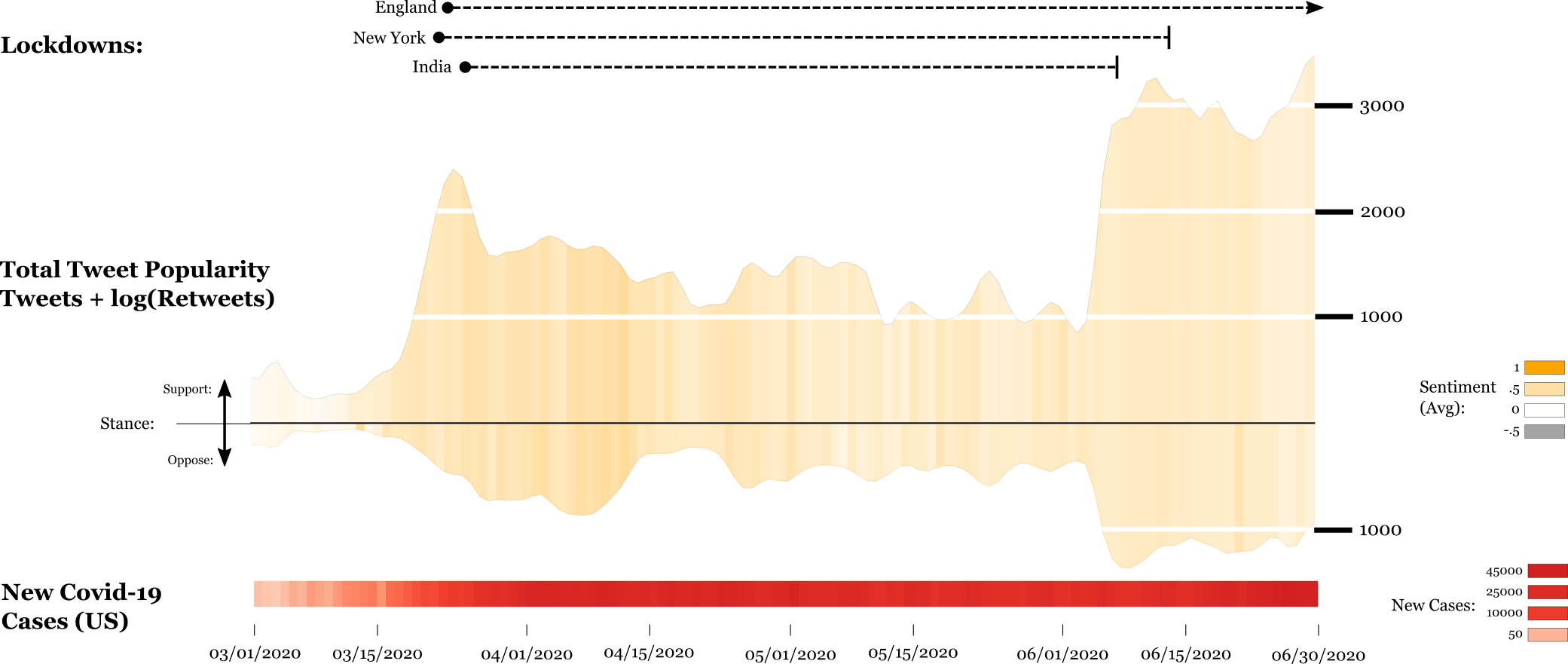}
        \caption{Timeline of major lockdowns, number of tweets weighted by number of retweets and COVID-19 cases. Tweets above and below the center axis represent tweets for and against SAH orders, respectively. Color encodes average daily sentiment score of tweets. %
        }
    \label{fig:alltweets}
    \vspace{-10pt}
\end{figure*}

To gain additional insight into the relationship between moral frames and stance, we performed a $\chi^2$ post-hoc analysis, and a regression analysis on the global data. Table~\ref{table:mf-stance} shows the contingency table for our 12 moral frames and 3 stances. A $\chi^2$ test confirmed  a strong  association between the two variables ($\chi^2 = 7,640.089, \: p<0.001$). 

We further investigate  which  moral frames indicate which specific stance using post-hoc tests based on the adjusted residuals for each cell (shown in parentheses in Table~\ref{table:mf-stance}). Here we find the largest differences between expected and observed counts, relative to sample size. %
According to \cite{agresti2019,sharpe2015chi}, adjusted residuals  with an absolute value greater than~3 for tables with many cells indicate a significant deviation from the expected value. 
Hence,  positive adjusted residuals greater than~3 indicate leaning of a moral frame more towards the corresponding stance than expected by chance. On the other hand, a negative adjusted residual indicates an association of a moral frame with the corresponding stance, lower than expected by chance. 

The adjusted residual values paint a rather striking picture. %
As concerns positive stance, Care (in bold in Table~\ref{table:mf-stance}) is the only moral frame whose residual  is large and positive, and hence, is associated with positive stance (all other residuals in the \emph{positive stance} column are negative, with a couple - Betrayal and Subversion - being of too small magnitude). The picture is reversed for negative stance:  only Care has a large negative residual in this column, all the other frames have large positive residuals (other than Subversion), confirming their association with negative stance. The four we have underlined  in Table~\ref{table:mf-stance} - Harm, Purity, Freedom and Oppression -  are the ones more strongly associated with  negative stance. As concerns the undecided stance, only two frames,  Harm and Freedom, have large positive residuals; hence, differently from the other frames, Harm and Freedom   are significantly associated with two stances, against or undecided. 

We also report the values for predicting stance using moral frames, retweets, and sentiment using a logistic regression model.
Of the moral frames, only Care was positively correlated with supporting SAH tweets stance relative to average ($\beta=0.149$, $p < .01$), confirming the finding obtained via the adjusted residual analysis. 
Harm ($\beta=-0.06$, $p < .01$), Fairness ($\beta=-0.09$), Injustice ($\beta=-0.13$), Freedom ($\beta=-0.1$), Oppression ($\beta=-0.1$), Purity ($\beta=-0.09$), and Degradation ($\beta=-0.05$) were all significantly correlated with opposition to SAH ($p < .001$), relative to the average.  Again, this agrees to a large degree with the adjusted residual analysis that had found all these moral frames to be associated with negative stance.

We next analyze the annotated \emph{US-SAH} corpus with two goals in mind. First, to understand whether the findings of the \emph{SAH-REL} corpus hold in the nation where most of the geo-tagged tweets originate, and second, to understand whether they provide additional insights with respect to the partisanship of views.

To a large extent, the findings hold as discussed in the $\chi^2$ analysis of the global dataset--$63\%$ of the tweets that supported SAH invoked appeals to Care as moral justification. Affirming the potency of appeals to Care, tweets using this justification were also most likely to be retweeted. Thus, $26\%$ of tweets invoked cautionary appeals to the prevention of Harm to others. Loyalty to others and deference to the expertise of Authority (largely medical) tied for third place at about $13\%$ each.

Those opposed to SAH invoked appeals to the vice or virtue binaries of Freedom, at over $54\%$, the most common moral justification for resistance. Predictably, most of these—about $21\%$--referenced Oppression. The remaining $33\%$ cited appeals to the virtue of Freedom. The second most cited frame was Harm ($34\%$), largely in reference to the economy. As Care tweets were most likely to be retweeted for supporters, Oppression tweets were most likely to be retweeted for those who opposed SAH orders. 

The vast majority of U.S. stay-at-home tweets originated from urban areas voting Democrat. Tweets containing a Freedom or Subversion moral frame were roughly equally distributed between Democrat and Republican voting areas. Different moral frames have similar geospatial distribution patterns across the US, with the exception of less frequent frames. Figure \ref{fig:geothreeframes} shows three such patterns for the Care, Oppression, and Degradation frames. The Degradation moral frame is extremely localized geographically in a small number (six) of non-adjacent counties such as Pecos County, Texas, and Leon County, Florida.

Because the \emph{SAH-REL} corpus was automatically labelled by our moral frame and stance classifiers, the results may be affected by the mistakes these classifiers make. Hence, we also ran $\chi^2$ and the adjusted residual analysis on the \emph{US-SAH} dataset, which has gold standard annotations as concerns both moral frames and stance. In that residual analysis, we found that Care was most strongly associated with a positive stance, %
whereas Freedom and Oppression in the \emph{US-SAH} dataset were the ones most strongly associated with negative stance, as was the case in the \emph{SAH-REL} corpus. In addition, Harm, Subversion and Injustice also were associated with negative stance. As stated earlier, Authority was likely to be associated with medical expertise as with that of governing officials.

\begin{figure}[h]
    \centering
     \includegraphics[width=\columnwidth]{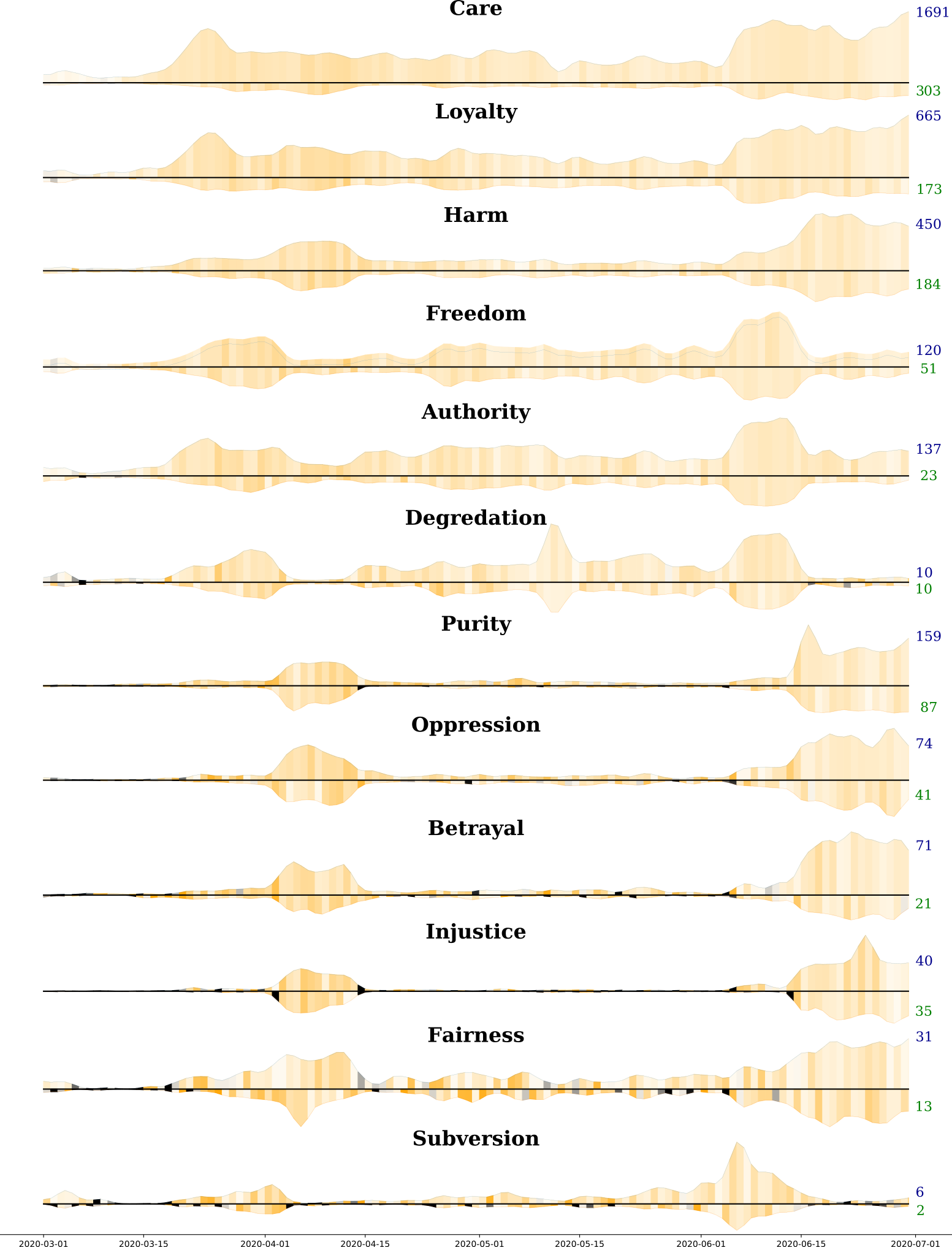}
        \caption{Ordering of frames by popularity over time based on tweets that contain them, colored by daily average sentiment. %
        }
        \label{fig:sparklines}
        \vspace{-10pt}
\end{figure}

\begin{figure*}[h]
    \centering
     \includegraphics[width=\textwidth]{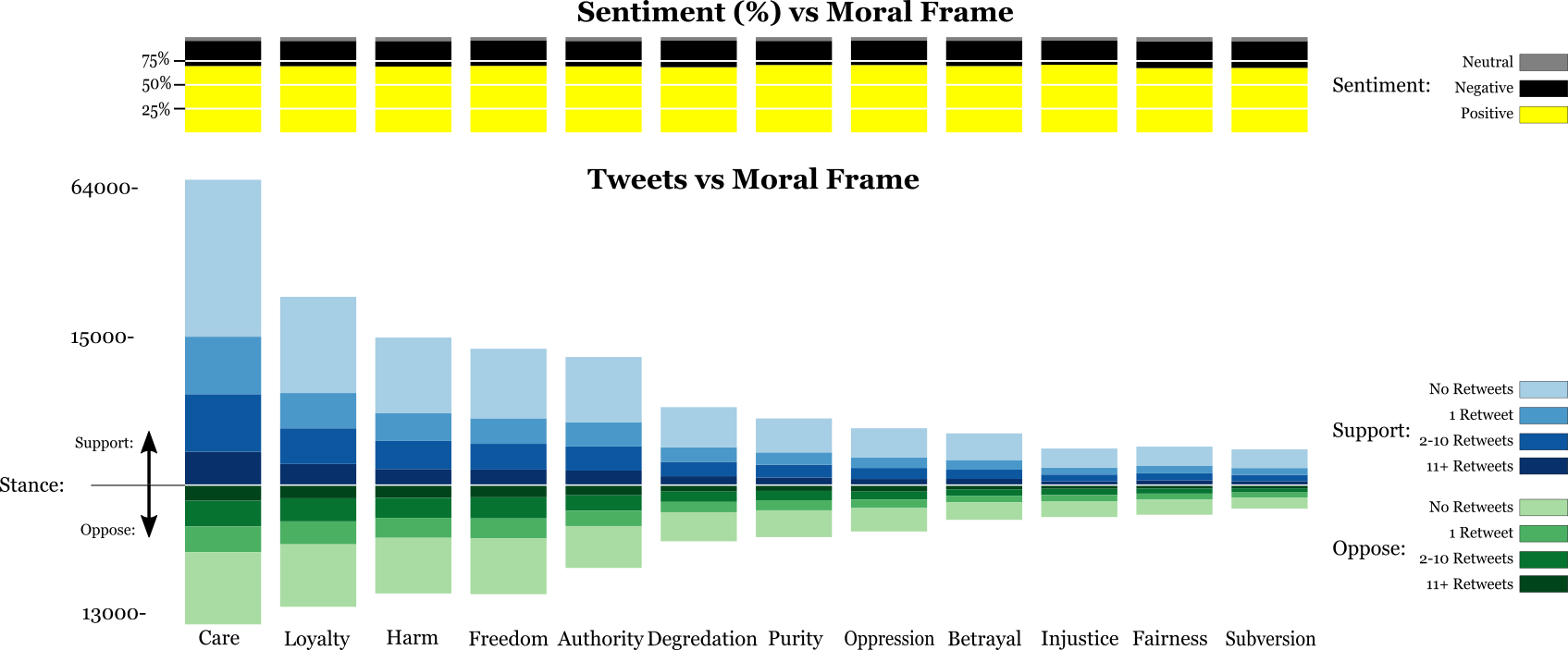}
        \caption{Moral frames reflected in SAH tweets and their respective stance distribution (bottom, log scale) and sentiment (top). %
        Tweets are predominantly in support of SAH, with Care being the most in-support ($82.7\%$), Oppression being the least in-support ($60.9\%$).}  %
        \label{fig:frames}
        \vspace{-10pt}
\end{figure*}

\subsection{Volume of tweets supporting and opposing SAH corresponds to real-world events}

To assess the stance of stay-at-home tweets over time, we examine the stance expressed in the observed tweets in the context of COVID-19 case counts.  To understand the spread of each tweet $i$, we calculate its overall virality based on its retweets. We use the following virality formula which reflects the scale of spread: $virality_i = 1+ ln(retweets_i + 1)$.  
Then the overall $virality$ of tweets on a given day is $\sum_{i} virality_i$.

Fig.~\ref{fig:alltweets} shows the total virality of SAH-related tweets over time via a set of time-based visual  encodings~\cite{marai2018precision, ma2017prodigen}. The majority ($73\%$) of stay-at-home tweets are in support of stay-at-home orders, with a large spike in the number of tweets, approximately between March 23rd and April 15\ts{th}, corresponding with the first month of stay-at-home orders being issued in several areas. The period of activity during March is characterized by lower overall COVID rates (shown in the bottom timeline chart~\cite{luciani2014large,luciani2014fixingtim}), more tweets supporting SAH orders, and a higher average sentiment, compared to later months.  The most viral tweet occurs at the beginning of this spike on March 4\ts{th} with 128,868 retweets. Its moral components invoke Care and (medical) Authority: "\textit{Protect yourself and your community from coronavirus with common sense precautions: wash your hands, stay home when sick and listen to the @CDCgov and local health authorities. Save the masks for health care workers. Let’s stay calm, listen to the experts, and follow the science.}" 

Tweets in support of SAH orders decreased between April 15\ts{th} and early June, although the number of tweets opposed to SAH orders remained relatively constant during this time. This may have been due to a lull in the initial interest, as well as the onset of summer, and the greater focus on the Black-Lives matter movement in Western nations. %
A notable, sustained spike in Twitter activity occurs on both sides of the issue at the beginning of June, starting at around June 7\ts{th}. This activity corresponds to the time that reopening was being discussed or implemented in several nations. This included the end of India's first national lockdown, announced on June 1st. One viral tweet ($387$ retweets) from Mumbai pleaded "\textit{Dear Mumbaikars, The worst is NOT over. Please stop acting like it has and please stay home as much as you can.\#StaySafe.}"  A spike in activity may also be attributed to a suspected rise in cases during Memorial day and the international George Floyd protests, with one popular tweet from California saying "\textit{This is starting to make the rounds. Don’t be fooled. COVID takes 1-2 weeks." The latest spike is from those Memorial Day dopes. Keep fighting! Stay safe \& if you're immuno-compromised or feeling ill STAY HOME! There are lots of ways to show your support!}." 
The most popular tweet during this time, with $17,198$ retweets, echoed a similar sentiment with a new report on COVID-related deaths: "\textit{A least 6,000 people have died from COVID-19 in June in the U.S. as the pandemic continues to rage. We remind everyone to please continue social distancing, mask-wearing, and safe practices. Check up on your elders and self-quarantine. It is only going to get worse from here on.}"

A small peak in anti-SAH tweets on June 9\ts{th} can be seen in Fig.~\ref{fig:alltweets}, which appears to be in response to a study by the World Health Organization, with one tweet with $663$ retweets from New York saying "\textit{Now the research from WHO is showing that the risk of asymptomatic people transmitting coronavirus is extremely rare. If that is true, that means we do not need social distancing...}", and another tweet with $550$ retweets saying "\textit{FFS. The World Health Organization strikes again. If we'd known this at the start of the pandemic, social distancing/lockdown measures could have been much less severe than those implemented.}"

\subsection{The relationship between moral frames, sentiment and virality changes over time}

To test our second hypothesis, we look at the relationship between moral frames found in \emph{SAH-REL} tweets, their sentiment and virality. 
Fig.~\ref{fig:sparklines} shows that 
Care, Loyalty, Freedom, and Authority are the predominant moral frames in the beginning of the pandemic. In contrast, Purity, Subversion, Degradation, and Betrayal become relatively more common in April when some of the first protests against SAH orders started and then again June when many of the lockdowns were lifted.
Analysis of the overall moral frame presence in tweets with stay-at-home attitude between March and June $2020$ shows that the Care frame is the most common moral frame ($40.1\%$).  
Next are Loyalty ($17.7\%$), Harm ($11.8\%$) and Freedom ($10.7\%$) (shown in compact visual form~\cite{aurisano2015bactogenie, ma2018rembrain} in Fig.~\ref{fig:frames}). 

In terms of sentiment (Fig.~\ref{fig:frames} top), the tweets with the highest portion of positive tweets are Injustice ($71\%$ positive), Purity ($70.9\%$ positive), and Oppression ($70.7\%$ positive). In contrast, the tweets with predominantly negative sentiment were Subversion ($28\%$ negative), Fairness ($27.9\%$ negative), Degradation ($27.7\%$ negative), and Harm ($26.8\%$ negative).

The majority ($86\%$) of tweets had few retweets (0-1), while only $3.3\%$ of tweets had more than $10$ retweets. To identify if moral frames, sentiment, or stance are correlated with tweet virality, we built multivariate regression to predict virality, given by $ln(retweets + 1)$, of each tweet, as well as stance, using moral frames, stance, and sentiment score. %
We report the regression coefficients ($\beta$), which correspond to the average expected increase in popularity when a tweet includes a given frame, relative to the average.
Sentiment is positively correlated with popularity ($\beta=0.047$, $p =0.003$), while moral frames do not have any statistically significant correlations. Freedom is slightly positively correlated with popularity ($\beta=0.0047$, $p = 0.536$). All the other moral frames have slightly negative correlation with popularity, with Injustice ($\beta=-0.037$, $p =0.059$), Subversion ($\beta=-0.038$, $p =0.081$) and Degradation ($\beta=-0.019$, $p =0.087$) having the most pronounced negative coefficients with low p-values.
P-values are calculated using a t-test to measure if coefficients are significantly different than zero. All models were built using the python \emph{statsmodels} package~\cite{seabold2010statsmodels}.

\section{Discussion}

The reactions of Twitter users sent to their networks of followers at the earliest stage of the pandemic preview the moral foundations of the divisions that hardened into ideological positions the following year. Though unrepresentative of the population in general, the opinions voiced on this platform offered a glimpse into trends that would divide citizens from each other. Research shows that though Twitter users may be unrepresentative of the population, they begin discussions that may persuade others, in part because they tend to be more politically active~\cite{wojcik-prc19}.%

Studying opinion accompanied by moral judgment is also warranted by its distinction from non-attitudes elicited on surveys. Moral judgment is a mark of commitment. It can motivate people to shame and punish wrongdoers and thereby to lead to cooperation and solidarity by denouncing those who flout the rules. But it also risks conflict by dehumanizing others and escalating into feuds~\cite{Crockett2017}. In the case of the covid pandemic, a moral contest of wills ensued that pitted supporters of medical expertise against those who prized freedom of choice. Medical advisors and health agencies were drawn into a polarized discourse that reflected political divisions rather than the authority of scientific expertise. As a result, mortality increased as a function of low vaccination rates, mask wearing, and other measures intended to slow the spread of the virus.

Our findings show overwhelming support for government mandates among those voicing an opinion on Twitter. Buttressed by concern for others and anger at those who disobeyed, supporters defined a line of moral conduct that would divide them from those who prized individual freedom over collective fate. The protests in late spring emboldened opponents who chafed against restrictions, mocking the hypocrisy and subversion of protesters who left their homes to score political points. Group loyalty emerged as a potential common ground, but partisan polarization reinforced by moral judgement used loyalty to maintain boundaries.

The significance of our study highlights the value of studying the germination of issue positions posted on social media where moral judgment is a valued currency of exchange. Future research on those issues most likely to elicit such judgment will supplement our knowledge of public opinion that has matured sufficiently to be analyzed by gold standard survey research measures.

\commentout{
\color{blue}
zf:We need to make these words consistent:\\
1) COVID-19 or COVID-19 - ez:should be COVID-19 \checkmark\\
2) SAH or stay-at-home - ez:should be stay-at-home  \checkmark \\
3) million or M - ez:should be million  \checkmark\\
4) US or U.S. - ez:should be U.S. \checkmark\\
5) 1805 or 1779 relevant tweets\\
6) attitude or stance - attitude until table 2, stance after.\\
7) Con or against - - ez:should be against \\
8) present tense vs. past tense - present tense
\color{black}
}

\section{Acknowledgements}
This research was partly supported by the National Science Foundation under grant No. 2031095.

\bibliographystyle{IEEEtran}
\bibliography{IEEEabrv,references}

\begin{thebibliography}{10}
\providecommand{\url}[1]{#1}
\csname url@samestyle\endcsname
\providecommand{\newblock}{\relax}
\providecommand{\bibinfo}[2]{#2}
\providecommand{\BIBentrySTDinterwordspacing}{\spaceskip=0pt\relax}
\providecommand{\BIBentryALTinterwordstretchfactor}{4}
\providecommand{\BIBentryALTinterwordspacing}{\spaceskip=\fontdimen2\font plus
\BIBentryALTinterwordstretchfactor\fontdimen3\font minus
  \fontdimen4\font\relax}
\providecommand{\BIBforeignlanguage}[2]{{%
\expandafter\ifx\csname l@#1\endcsname\relax
\typeout{** WARNING: IEEEtran.bst: No hyphenation pattern has been}%
\typeout{** loaded for the language `#1'. Using the pattern for}%
\typeout{** the default language instead.}%
\else
\language=\csname l@#1\endcsname
\fi
#2}}
\providecommand{\BIBdecl}{\relax}
\BIBdecl

\bibitem{Yin-adma20}
H.~Yin, S.~Yang, and J.~Li, ``Detecting topic and sentiment dynamics due to
  covid-19 pandemic using social media,'' in \emph{Advanced Data Mining and
  Applications}.\hskip 1em plus 0.5em minus 0.4em\relax Springer, 2020, pp.
  610--623.

\bibitem{li-po21}
L.~Li, A.~Erfani, Y.~Wang, and Q.~Cui, ``Anatomy into the battle of supporting
  or opposing reopening amid the covid-19 pandemic on twitter: A temporal and
  spatial analysis,'' \emph{Plos one}, vol.~16, no.~7, p. e0254359, 2021.

\bibitem{Bloom2013}
P.~B.-N. Bloom, ``The public’s compass: Moral conviction and political
  attitudes,'' \emph{American Politics Research}, vol.~41, no.~6, pp. 937--964,
  2013.

\bibitem{Scott2010}
G.~S. Morgan, L.~J. Skitka, and D.~C. Wisneski, ``Moral and religious
  convictions and intentions to vote in the 2008 presidential election,''
  \emph{Analyses of Social Issues and Public Policy}, vol.~10, no.~1, pp.
  307--320, 2010.

\bibitem{Skitka2008}
L.~J. Skitka and C.~W. Bauman, ``Moral conviction and political engagement,''
  \emph{Political Psychology}, vol.~29, no.~1, pp. 29--54, 2008.

\bibitem{Aramovich2012}
N.~P. Aramovich, B.~L. Lytle, and L.~J. Skitka, ``Opposing torture: Moral
  conviction and resistance to majority influence,'' \emph{Social Influence},
  vol.~7, no.~1, pp. 21--34, 2012.

\bibitem{Hornsey2003}
M.~J. Hornsey, L.~Majkut, D.~J. Terry, and B.~M. McKimmie, ``On being loud and
  proud: Non-conformity and counter-conformity to group norms,'' \emph{British
  Journal of Social Psychology}, vol.~42, no.~3, pp. 319--335, 2003.

\bibitem{valenzuela2017}
S.~Valenzuela, M.~Piña, and J.~Ramírez, ``{Behavioral Effects of Framing on
  Social Media Users: How Conflict, Economic, Human Interest, and Morality
  Frames Drive News Sharing},'' \emph{Journal of Communication}, vol.~67,
  no.~5, pp. 803--826, 2017.

\bibitem{Crockett2017}
M.~J. Crockett, ``Moral outrage in the digital age,'' \emph{Nature Human
  Behaviour}, vol.~1, no.~11, pp. 769--771, Nov 2017.

\bibitem{graham-jpsp09}
J.~Graham, J.~Haidt, and B.~Nosek, ``Liberals and conservatives rely on
  different sets of moral foundations,'' \emph{Journal of personality and
  social psychology}, vol.~96, pp. 1029--46, 06 2009.

\bibitem{Entman93}
R.~M. Entman, ``\BIBforeignlanguage{eng}{Framing: Toward clarification of a
  fractured paradigm},'' \emph{\BIBforeignlanguage{eng}{Journal of
  communication}}, vol.~43, no.~4, pp. 51--58, 1993.

\bibitem{chong2007}
D.~Chong and J.~N. Druckman, ``Framing theory,'' \emph{Annual Review of
  Political Science}, vol.~10, no.~1, pp. 103--126, 2007.

\bibitem{garten-ijcai16}
J.~Garten, R.~Boghrati, J.~Hoover, K.~M. Johnson, and M.~Dehghani, ``Morality
  between the lines : Detecting moral sentiment in text,'' in \emph{IJCAI},
  2016.

\bibitem{wojcik-prc19}
S.~Wojcik and A.~Hughes, ``Sizing up twitter users,'' \emph{Pew Research
  Center}, vol.~24, 2019.

\bibitem{grossman-nac20}
G.~Grossman, S.~Kim, J.~M. Rexer, and H.~Thirumurthy, ``Political partisanship
  influences behavioral responses to governors{\textquoteright} recommendations
  for covid-19 prevention in the united states,'' \emph{Proceedings of the
  National Academy of Sciences}, vol. 117, no.~39, pp. 24\,144--24\,153, 2020.

\bibitem{allcott-jpe20}
H.~Allcott, L.~Boxell, J.~Conway, M.~Gentzkow, M.~Thaler, and D.~Yang,
  ``Polarization and public health: Partisan differences in social distancing
  during the coronavirus pandemic,'' \emph{Journal of Public Economics}, vol.
  191, p. 104254, 2020.

\bibitem{painter-jebo21}
M.~Painter and T.~Qiu, ``Political beliefs affect compliance with government
  mandates,'' \emph{Journal of Economic Behavior \& Organization}, vol. 185,
  pp. 688--701, 2021.

\bibitem{roozenbeek-jrc20}
J.~Roozenbeek, C.~R. Schneider, S.~Dryhurst, J.~Kerr, A.~L. Freeman,
  G.~Recchia, A.~M. Van Der~Bles, and S.~Van Der~Linden, ``Susceptibility to
  misinformation about covid-19 around the world,'' \emph{Royal Society open
  science}, vol.~7, no.~10, p. 201199, 2020.

\bibitem{johnson-nature20}
N.~F. Johnson, N.~Vel{\'a}squez, N.~J. Restrepo, R.~Leahy, N.~Gabriel,
  S.~El~Oud, M.~Zheng, P.~Manrique, S.~Wuchty, and Y.~Lupu, ``The online
  competition between pro-and anti-vaccination views,'' \emph{Nature}, vol.
  582, no. 7811, pp. 230--233, 2020.

\bibitem{wen-ying-hc20}
W.-Y.~S. Chou and A.~Budenz, ``Considering emotion in covid-19 vaccine
  communication: Addressing vaccine hesitancy and fostering vaccine
  confidence,'' \emph{Health Communication}, vol.~35, no.~14, pp. 1718--1722,
  2020.

\bibitem{karami-aist20}
A.~Karami and M.~Anderson, ``Social media and covid-19: Characterizing
  anti-quarantine comments on twitter,'' \emph{Proceedings of the Association
  for Information Science and Technology}, vol.~57, no.~1, p. e349, 2020.

\bibitem{aLdayel-ipm21}
A.~ALDayel and W.~Magdy, ``Stance detection on social media: State of the art
  and trends,'' \emph{Information Processing \& Management}, vol.~58, no.~4, p.
  102597, 2021.

\bibitem{sobhani-lcs16}
P.~Sobhani, S.~Mohammad, and S.~Kiritchenko, ``Detecting stance in tweets and
  analyzing its interaction with sentiment,'' in \emph{Proceedings of the Fifth
  Joint Conference on Lexical and Computational Semantics}.\hskip 1em plus
  0.5em minus 0.4em\relax ACL, Aug. 2016, pp. 159--169.

\bibitem{darwish-icwsm20}
K.~Darwish, P.~Stefanov, M.~Aupetit, and P.~Nakov, ``Unsupervised user stance
  detection on twitter,'' in \emph{ICWSM}, vol.~14, 2020, pp. 141--152.

\bibitem{mohammad-semeval16}
S.~Mohammad, S.~Kiritchenko, P.~Sobhani, X.~Zhu, and C.~Cherry,
  ``{S}em{E}val-2016 task 6: Detecting stance in tweets,'' in \emph{Proceedings
  of the 10th International Workshop on Semantic Evaluation
  ({S}em{E}val-2016)}.\hskip 1em plus 0.5em minus 0.4em\relax ACL, Jun. 2016,
  pp. 31--41.

\bibitem{mitra-icwsm17}
T.~Mitra, S.~Counts, and J.~Pennebaker, ``Understanding anti-vaccination
  attitudes in social media,'' \emph{ICWSM}, vol.~10, no.~1, Mar. 2016.

\bibitem{hassan-emnlp10}
A.~Hassan, V.~Qazvinian, and D.~Radev, ``What{'}s with the attitude?
  identifying sentences with attitude in online discussions,'' in
  \emph{EMNLP}.\hskip 1em plus 0.5em minus 0.4em\relax ACL, Oct. 2010.

\bibitem{valenzuela-joc17}
S.~Valenzuela, M.~Piña, and J.~Ramírez, ``Behavioral effects of framing on
  social media users: How conflict, economic, human interest, and morality
  frames drive news sharing,'' \emph{Journal of Communication}, vol.~67, no.~5,
  pp. 803--826, 2017.

\bibitem{card-emnlp16}
D.~Card, J.~Gross, A.~Boydstun, and N.~A. Smith, ``Analyzing framing through
  the casts of characters in the news,'' in \emph{EMNLP}.\hskip 1em plus 0.5em
  minus 0.4em\relax Austin, Texas: ACL, Nov. 2016, pp. 1410--1420.

\bibitem{shahid-acl20}
U.~Shahid, B.~Di~Eugenio, A.~Rojecki, and E.~Zheleva, ``Detecting and
  understanding moral biases in news,'' in \emph{Proceedings of the First Joint
  Workshop on Narrative Understanding, Storylines, and Events}.\hskip 1em plus
  0.5em minus 0.4em\relax ACL, 2020, pp. 120--125.

\bibitem{johnson-acl17}
K.~Johnson, D.~Jin, and D.~Goldwasser, ``Leveraging behavioral and social
  information for weakly supervised collective classification of political
  discourse on {T}witter,'' in \emph{ACL}, Jul. 2017, pp. 741--752.

\bibitem{hoover-spps20}
J.~Hoover, G.~Portillo-Wightman, L.~Yeh, S.~Havaldar, A.~M. Davani, Y.~Lin,
  B.~Kennedy, M.~Atari, Z.~Kamel, M.~Mendlen, G.~Moreno, C.~Park, T.~E. Chang,
  J.~Chin, C.~Leong, J.~Y. Leung, A.~Mirinjian, and M.~Dehghani, ``Moral
  foundations twitter corpus: A collection of 35k tweets annotated for moral
  sentiment,'' \emph{Social Psychological and Personality Science}, vol.~11,
  no.~8, pp. 1057--1071, 2020.

\bibitem{kaur-bigdata16}
R.~{Kaur} and K.~{Sasahara}, ``Quantifying moral foundations from various
  topics on twitter conversations,'' in \emph{2016 IEEE International
  Conference on Big Data}, 2016, pp. 2505--2512.

\bibitem{eisenstein-naacl2013}
J.~Eisenstein, ``What to do about bad language on the internet,'' in
  \emph{Proceedings of the 2013 conference of the North American Chapter of the
  Association for Computational Linguistics: Human language technologies},
  2013, pp. 359--369.

\bibitem{sarker-norm2017}
A.~Sarker, ``A customizable pipeline for social media text normalization,''
  \emph{Social Network Analysis and Mining}, vol.~7, no.~1, pp. 1--13, 2017.

\bibitem{skaik-acm17}
R.~Skaik and D.~Inkpen, ``Using social media for mental health surveillance: A
  review,'' \emph{ACM Comput. Surv.}, vol.~53, no.~6, dec 2020.

\bibitem{chen-jmir20}
E.~Chen, K.~Lerman, and E.~Ferrara, ``{Tracking Social Media Discourse About
  the COVID-19 Pandemic: Development of a Public Coronavirus Twitter Data
  Set},'' p. e19273, 2020.

\bibitem{morstatter-aaai13}
F.~Morstatter, J.~Pfeffer, H.~Liu, and K.~Carley, ``Is the sample good enough?
  comparing data from twitter's streaming api with twitter's firehose,'' in
  \emph{ICWSM}, vol.~7, no.~1, 2013.

\bibitem{blei-jmlr}
D.~M. Blei, A.~Y. Ng, and M.~I. Jordan, ``{Latent Dirichlet Allocation},''
  \emph{Journal of Machine Learning Research}, vol.~3, pp. 993--1022, 2003.

\bibitem{miller-princeton}
G.~Miller, R.~Beckwith, C.~Fellbaum, D.~Gross, and K.~J. Miller,
  ``{Introduction to WordNet: An On-line Lexical Database},''
  \emph{International Journal of Lexicography}, vol.~3, pp. 235--244, 1990.

\bibitem{mikolov-acm}
T.~Mikolov, I.~Sutskever, K.~Chen, G.~Corrado, and J.~Dean, ``{Distributed
  representations of words and phrases and their compositionality},''
  \emph{Proceedings of the 26th International Conference on Neural Information
  Processing Systems - Volume 2}, pp. 3111--3119, Dec 2013.

\bibitem{haidt2012}
J.~Haidt, \emph{The righteous mind: Why good people are divided by politics and
  religion}.\hskip 1em plus 0.5em minus 0.4em\relax Vintage, 2012.

\bibitem{krippen80}
K.~Krippendorff, \emph{Content Analysis: an Introduction to its
  Methodology}.\hskip 1em plus 0.5em minus 0.4em\relax Beverly Hills, CA: Sage
  Publications, 1980.

\bibitem{hutto-aaai14}
C.~Hutto and E.~Gilbert, ``Vader: A parsimonious rule-based model for sentiment
  analysis of social media text,'' in \emph{ICWSM}, vol.~8, no.~1, 2014.

\bibitem{snorkel}
A.~Ratner, S.~H. Bach, H.~Ehrenberg, J.~Fries, S.~Wu, and C.~R\'{e}, ``Snorkel:
  Rapid training data creation with weak supervision,'' \emph{Proc. VLDB
  Endow.}, vol.~11, no.~3, p. 269–282, Nov. 2017.

\bibitem{adasyn}
H.~He, Y.~Bai, E.~A. Garcia, and S.~Li, ``Adasyn: Adaptive synthetic sampling
  approach for imbalanced learning,'' in \emph{2008 IEEE International Joint
  Conference on Neural Networks (IEEE World Congress on Computational
  Intelligence)}, 2008, pp. 1322--1328.

\bibitem{pennington-etal-2014-glove}
J.~Pennington, R.~Socher, and C.~Manning, ``{G}lo{V}e: Global vectors for word
  representation,'' in \emph{EMNLP}.\hskip 1em plus 0.5em minus 0.4em\relax
  ACL, Oct. 2014, pp. 1532--1543.

\bibitem{Graham2013}
J.~Graham, J.~Haidt, S.~Koleva, M.~Motyl, R.~Iyer, S.~P. Wojcik, and P.~H.
  Ditto, ``Moral foundations theory: The pragmatic validity of moral
  pluralism,'' in \emph{Advances in experimental social psychology}.\hskip 1em
  plus 0.5em minus 0.4em\relax Elsevier, 2013, vol.~47, pp. 55--130.

\bibitem{trelles2019visual}
J.~Trelles, D.~Lee, S.~Derrible, and G.~E. Marai, ``{Visual Analysis of a Smart
  City’s Energy Consumption},'' \emph{Multimodal Technologies and
  Interaction}, vol.~3, no.~2, p.~30, 2019.

\bibitem{agresti2019}
A.~Agresti, \emph{An introduction to categorical data analysis}, 3rd~ed.\hskip
  1em plus 0.5em minus 0.4em\relax John Wiley \& Sons, 2019.

\bibitem{sharpe2015chi}
D.~Sharpe, ``Chi-square test is statistically significant: Now what?''
  \emph{Practical Assessment, Research, and Evaluation}, vol.~20, no.~1, 2015,
  article 8.

\bibitem{marai2018precision}
G.~E. {Marai}, C.~{Ma}, A.~T. {Burks}, F.~{Pellolio} \emph{et~al.},
  ``{Precision Risk Analysis of Cancer Therapy with Interactive Nomograms and
  Survival Plots},'' \emph{IEEE Trans. Vis. Comp. Graph.}, vol.~25, no.~4, p.
  1732–1745, 2019.

\bibitem{ma2017prodigen}
C.~Ma, T.~Luciani \emph{et~al.}, ``{PRODIGEN: visualizing the probability
  landscape of stochastic gene regulatory networks in state and time space},''
  \emph{BMC Bioinform.}, vol.~18, no.~2, pp. 1--14, 2017.

\bibitem{luciani2014large}
T.~Luciani, B.~Cherinka, D.~Oliphant \emph{et~al.}, ``Large-scale overlays and
  trends: Visually mining, panning and zooming the observable universe,''
  \emph{IEEE Trans. Vis. Comp. Graph.}, vol.~20, no.~7, pp. 1048--1061, 2014.

\bibitem{luciani2014fixingtim}
T.~Luciani, J.~Wenskovitch \emph{et~al.}, ``{FixingTIM}: interactive
  exploration of sequence and structural data to identify functional mutations
  in protein families,'' in \emph{BMC proceedings}, vol.~8, no.~2.\hskip 1em
  plus 0.5em minus 0.4em\relax Springer, 2014, pp. 1--9.

\bibitem{aurisano2015bactogenie}
J.~Aurisano, K.~Reda \emph{et~al.}, ``{BactoGeNIE}: a large-scale comparative
  genome visualization for big displays,'' \emph{BMC Bioinformatics}, vol.~16,
  no.~11, pp. 1--14, 2015.

\bibitem{ma2018rembrain}
C.~Ma, F.~Pellolio, D.~A. Llano \emph{et~al.}, ``{Rembrain: Exploring dynamic
  biospatial networks with mosaic matrices and mirror glyphs},''
  \emph{Electronic Imaging}, vol. 2018, no.~1, pp. 060\,404--1, 2018.

\bibitem{seabold2010statsmodels}
S.~Seabold and J.~Perktold, ``statsmodels: Econometric and statistical modeling
  with python,'' in \emph{9th Python in Science Conference}, 2010.

\end{thebibliography}

\end{document}